\documentclass[11pt]{article}
\usepackage{hyperref}
\pdfoutput=1

\begin{document}
\title{Electro-Coalescence Fireworks}
\author{H. Aryafar and H. P. Kavehpour \\
\\\vspace{6pt} Henry Samueli School of Engineering and Applied Science, \\ University of California, Los Angeles, CA 90095, USA}

\maketitle

\begin{abstract}
Electro-coalescence is the application of an electric field onto coalescing fluid bodies. The following fluid dynamics videos show a droplet coalescing into a fluid bath while embedded into a viscous medium and subject to a very high electric field. The concentration of electric stresses at the apex of the droplet cause it to break apart. The droplet is glycerol and the viscous medium is silicone oil. 
\href{http://ecommons.library.cornell.edu/bitstream/1813/13795/2/Aryafar_GFM_2009.mpg}{Full Resolution} 
and 
\href{http://ecommons.library.cornell.edu/bitstream/1813/13795/3/Aryafar_GFM_2009_low_res.mpg}{Low Resolution}.

\end{abstract}

\end{document}